\documentstyle[preprint,floats,prl,aps,epsf]{revtex}

\begin{document}

\preprint{\vbox{\hbox {December 1999} \hbox{IFP-778-UNC}
\hbox{VAND-TH-99-11}}}
%\twocolumn[\hsize\textwidth\columnwidth\hsize\csname@twocolumnfalse\endcsname

\draft
\title{Chiral Fermions and AdS/CFT Duality for a Nonabelian Orbifold}
\author{\bf Paul H. Frampton$^{(a)}$ and Thomas W. Kephart$^{(b)}$}
\address{(a)Department of Physics and Astronomy,\\
University of North Carolina, Chapel Hill, NC  27599.}
\address{(b)Department of Physics and Astronomy,\\
 Vanderbilt University, Nashville, TN 37325.}
\maketitle
\date{\today}

\begin{abstract}
We present what we believe is the minimal three-family $AdS/CFT$ model
compactified on a nonabelian orbifold $S^{5}/(Q\times Z_{3})$.
Nontrivial irreps of the discrete nonabelian group
$Q\times Z_{3}$ are identified with the $4$ of $SU(4)$ $R$
symmetry to break all supersymmetries, and the scalar content of the
model is sufficient to break the gauge symmetry to the standard model.
According to the conformality hypothesis the progenitor $SU(4)^3 \times SU(2)^{12}$
theory becomes conformally invariant at an infra-red fixed point of the
renormalization group.
\end{abstract}

\newpage

The concept of duality has always proven to be one of the most useful
tools in advancing theoretical physics, beginning with
the duality
between {\bf x} and {\bf p} in classical and quantum mechanics.
Further examples include wave-particle duality which is closely related
to {\bf x-p} duality, the duality
between {\bf E} and {\bf B} in Maxwell's equations,
the instanton solutions of self-dual Yang-Mills equations,
the Kramers-Wannier duality between low and high temperature
in a condensed-matter phase
transition, and dual resonance models which were the precursor to
string theory. More recently there is the duality between weak and
strong coupling
field theories and then between all the different superstring theories
that has led to a revolution in our understanding of strings.
Most recently, and equally profound, is the AdS/CFT duality
which is the subject of the present article. This AdS/CFT duality is
between
string theory compactified on Anti-de-Sitter space and Conformal Field
Theory.

Until very recently, the possibility of testing string theory seemed at
best remote. The advent of $AdS/CFT$s and large-scale string
compactification suggest this point of view may be too pessimistic,
since both could lead to $\sim 100TeV$ evidence for strings. With this
thought in mind, we are encouraged to build $AdS/CFT$ models with
realistic fermionic structure, and reduce to the standard model below
$\sim 1TeV$.

Using AdS/CFT duality, one arrives at a class of gauge field theories
of special recent interest. The simplest compactification of a
ten-dimensional
superstring on a product of
an AdS space with a five-dimensional spherical manifold leads to
an ${\cal N} = 4~SU(N)$ supersymmetric gauge theory, well known to be
conformally invariant\cite{mandelstam}. By replacing the manifold $S^5$
by an orbifold $S^5/\Gamma$ one arrives at less supersymmetries
corresponding to ${\cal N} = 2,~1 ~{\rm or}~ 0$ depending\cite{KS}
on whether $\Gamma \subset SU(2), ~~ SU(3),
~~{\rm or} \not\subset SU(3)$ respectively, where $\Gamma$
is in all cases a subgroup of $SU(4) \sim SO(6)$
the isometry of the $S^5$ manifold.

It was conjectured in \cite{maldacena} that such $SU(N)$ gauge theories
are conformal in the $N \rightarrow \infty$ limit. In \cite{F1} it was
conjectured that
at least a subset of the resultant nonsupersymmetric
${\cal N} = 0$ theories are conformal even for finite $N$. Some
first steps to check this idea were made in \cite{WS}.
Model-building based on abelian
$\Gamma$ was studied further in \cite{CV,F2,F3}, arriving in \cite{F3}
at an $SU(3)^7$ model based on $\Gamma = Z_7$
which has three families of chiral fermions,
a correct value for ${\rm sin}^2 \theta$ and a conformal scale $\sim
10$~~TeV.

The case of non-abelian orbifolds bases on non-abelian $\Gamma$ has
not previously been studied, partially due to the fact that it is
apparently somewhat more mathematically
sophisticated. However, we shall show here that it can be handled
equally
as systematically as the abelian case and leads to richer structures
and interesting results.

We consider all non-abelian discrete groups of order $g < 32$. These
are described in detail in \cite{books,FK}. There are exactly 45 such
non-abelian groups. Because the gauge group arrived at
by this construction\cite{CV}
is $\otimes_i SU(Nd_i)$ where $d_i$ are the dimensions of the
irreducible representations of $\Gamma$, one can expect to arrive
at models such as the Pati-Salam
$SU(4) \times SU(2) \times SU(2)$ type\cite{PS}
by choosing $N = 2$ and combining two singlets
and a doublet in the {\bf 4}
of $SU(4)$. Indeed we shall show that such an
accommodation of the standard model
is possible by using a non-abelian $\Gamma$.

The procedures for building a model within such a conformality approach
are:
(1) Choose $\Gamma$; (2) Choose a proper
embedding $\Gamma \subset SU(4)$ by assigning
the components of the {\bf 4} of $SU(4)$ to irreps of $\Gamma$,
while at the same time ensuring that the {\bf 6} of $SU(4)$ is real;
(3) Choose $N$, in the gauge group $\otimes_i SU(Nd_i)$. (4) Analyse the
patterns of spontaneous symmetry breaking.

In the present study we shall choose $N = 2$ and aim
at the gauge group $SU(4) \times SU(2) \times SU(2)$.
To obtain chiral fermions, it is necessary\cite{CV}
that the {\bf 4} of $SU(4)$
be complex ${\bf 4} \neq {\bf 4}^*$. Actually this condition is not
quite
sufficient to ensure chirality in the present case because
of the pseudoreality of $SU(2)$. We must ensure that the {\bf 4} is
not just pseudoreal.

This last condition means that many of our 45 candidates for $\Gamma$
do not lead to chiral fermions. For example, $\Gamma = Q_{2n} \subset
SU(2)$
has irreps of appropriate dimensionalities for our purpose
but it will not sustain chiral fermions under $SU(4)\times SU(2) \times
SU(2)$
because these
irreps are all, like $SU(2)$, pseudoreal.\footnote{Note that
were we using $N \geq 3$
then a pseudoreal {\bf 4} would give chiral fermions.}
Applying the rule that {\bf 4} must be
neither real nor pseudoreal
leaves a total of only 19 possible non-abelian discrete groups of
order $g \leq 31$. The smallest group which avoids pseudoreality
has order $g = 16$ but gives only two families. The
technical details
of our systematic search will be postponed
to a future publication. Here we shall present only the
simplest interesting non-abelian case
which has $g = 24$ and gives three chiral families in a
Pati-Salam-type model\cite{PS}.

Before proceeding to the details of
the specific $g = 24$ case, it is worth reminding the reader that
the Conformal Field Theory (CFT) that it exemplifies should
be free of all divergences,
even logarithmic ones, if the conformality conjecture is correct,
and be completely finite. Further
the theory is originating from a superstring theory
in a higher-dimension (ten) and contains gravity\cite{V,RS,GW}
by compactification of the
higher-dimensional graviton already contained in that superstring
theory. In
the CFT as we derive it, gravity is absent because we have not kept
these graviton modes - of course, their influence on high-energy
physics experiments is generally completely negligible unless the
compactification
scale is ``large''\cite{antoniadis}; here we shall neglect the effects
of
gravity.

To motivate our model it is instructive to comment on the choice of
$\Gamma$ and
on the
choice of embedding.

If we embed only four singlets of $\Gamma$ in the {\bf 4} of $SU(4)$
then this
has the effect of abelianizing $\Gamma$ and the gauge group obtained in
the
chiral sector of the
theory is $SU(N)^q$. These cases can be interesting but have already
been
studied\cite{CV,F2}.
Thus, we require at least one irrep of $\Gamma$ to have $d_i \geq 2$ in
the
embedding.

The only $\Gamma$ of order $g \leq 31$ with a {\bf 4} is $Z_5
\tilde{\times}
Z_4$ and this embedding
leads to a non-chiral theory. This leaves only embeddings with two
singlets and
a doublet,
a triplet and a singlet or two doublets.

The third of these choices leads to richer structures for
low order $\Gamma$. Concentrating on them shows that
of the chiral models possible, those from groups of low order
result in an insufficient number (below three) of
chiral families.

The first group that can lead to exactly three families occurs at
order $g = 24$ and is $\Gamma = Z_3 \times Q$ where $Q (\equiv Q_4)$
is the group of unit quarternions which is the smallest dicyclic group
$Q_{2n}$.

There are several potential models due to the different choices for the
{\bf 4}
of $SU(4)$ but only the case {\bf 4} = $(1\alpha, 1^{'}, 2\alpha)$ leads
to three families so let
us describe this in some detail:

Since $Q \times Z_3$ is a direct product group, we can write the irreps
as $R_i
\otimes \alpha^{a}$
where $R_i$ is a $Q$ irrep and $\alpha^{a}$ is a $Z_3$ irrep. We write
$Q$
irreps as $1,~1^{'},~1^{''},~
1^{'''},~2$ while the irreps of $Z_3$ are all singlets which we call
$\alpha, \alpha^2, \alpha^3 = 1$.
Thus $Q \times Z_3$ has fiveteen irreps in all and the gauge group will
be of Pati-Salam type for $N = 2$.

If we wish to break all supersymmetry, the {\bf 4} may not contain the
trivial
singlet of
$\Gamma$.
Due to permutational symmetry among the singlets
it is sufficiently general to choose {\bf 4} =
$(1\alpha^{a_1},~1^{'}\alpha^{a_2},~2\alpha^{a_3})$
with $a_1 \neq 0$.

To fix the $a_i$ we note that the scalar sector of the theory which is
generated
by the
{\bf 6} of $SU(4)$ can be used as a constraint since the {\bf 6} is
required to be real. This leads to
$a_1 + a_2 = - 2a_3 ({\rm mod}~3)$. Up to permutations in the chiral
fermion
sector the most
general choice is $a_1 = a_3= +1$ and $a_2 = 0$. Hence our choice of
embedding
is
\begin{equation}
{\bf 4} = (1\alpha,~1^{'},~2\alpha)
\label{embed}
\end{equation}
with
\begin{equation}
{\bf 6} = (1^{'}\alpha,~2\alpha,~2\alpha^{2},~1^{'}\alpha^{2})
\label{six}
\end{equation}
which is real as required.

We are now in a position to summarize the particle content of the
theory. The
fermions are given by
\begin{equation}
\sum_I~{\bf 4}\times R_I
\end{equation}
where the $R_I$ are all the irreps of $\Gamma = Q \times Z_3$. This is:
\[
\sum_{i=1}^{3} [(2_{1}\alpha^{i},2_{2}\alpha^{i})
+(2_{3}\alpha^{i},2_{4}\alpha^{i})+(2_{2}\alpha^{i},2_{1}\alpha^{i})
+(2_{4}\alpha^{i},2_{3}\alpha^{i})+(4\alpha^{i},\overline{4}\alpha^{i})] 
\]

\begin{equation}
+ \sum_{i=1}^{3} \sum_{a=1}^{4} [(2_{a}\alpha^{i},
2_{a}\alpha^{i+1})+(2_{a}\alpha^{i},4\alpha^{i+1})
+ (\bar{4}\alpha^{i},2_{a}\alpha^{i+1})]
\label{fermions}
\end{equation} 

 It is convenient to represent the chiral portions of these in a given
diagram (see Figure 1).

The scalars are given by
\begin{equation}
\sum_I~{\bf 6}\times R_I
\end{equation}
and are:
\[
\sum_{i=1}^{3} \sum_{j=1(j\neq i)}^{3} 
[(2_{1}\alpha^{i},2_{2}
\alpha^{j})+(2_{2}\alpha^{i}, 2_{1}\alpha^{j})+(2_{3}\alpha^{i},
2_{4}\alpha^{j})+(2_{4}\alpha^{i},2_{3}\alpha^{j})
+(2_{2}\alpha^{i},2_{1}\alpha^{i})+(2_{4}\alpha^{i},2_{3}\alpha^{i})] 
\]
\begin{equation}
+ \sum_{i=1}^{3} \sum_{j=1(j\neq i)}^{3} 
\{ \sum_{a=1}^{4}[(2_{a}\alpha^{i},4\alpha^{j})
+\bar{(4}\alpha^{i},2_{a}\alpha^{j} )]
+(4\alpha^{i}, \bar{4}\alpha ^{i}) \}
\label{scalars}
\end{equation}
which is easily checked to be real.

The gauge group $SU(4)^3 \times SU(2)^{12}$ with chiral fermions of
Eq.(\ref{fermions}) and scalars of Eq.(\ref{scalars}) 
is expected to acquire confromal invariance
at an infra-red fixed point of the  renormalization group, as discussed in \cite{F1}.

To begin our examination of the symmetry breaking we first
observe
that if we break the three $SU(4)$s to the totally diagonal $SU(4)$,
then chirality in the
fermionic sector is lost. To avoid this we break $SU_{1}(4)$ completely
and then break $SU_{\alpha }(4)\times SU_{\alpha ^{2}}(4)$ to
its diagonal subgroup $SU_{D}(4).$ The first of these steps can be
achieved with VEVs of the form $[(4_{1},2_{b}\alpha ^{k})+h.c.]$ where
we
are free to choose $b$, but $k$ must be $1$ or $2$ since there
are no $(4_{1},2_{b}\alpha ^{k=0})$ scalars. The second step requires an

$SU_{D}(4)$ singlet VEV from
($\overline{4}_{\alpha }$,4$_{\alpha^{2}})$ and/or
(4$_{\alpha }$, $\overline{4}_{\alpha ^{2}})$. Once we
make a choice for $b$ (we take $b=4$), the remaining chiral fermions
are,
in an intuitive notation:

\bigskip

\noindent $\ \sum_{a=1}^{3}\left[ (2_{a}\alpha \
,1,4_{D})+(1,2_{a}\alpha ^{-1},\overline{4_{D}})\right] $

\bigskip

\noindent which has the same content as as a three family Pati-Salam
model,
though with a separate $SU_{L}(2)\times SU_{R}(2)$ per family.

To further reduce the symmetry we must arrange to break to a single
$SU_{L}(2)$ and a single $SU_{R}(2).$ This is achieved by modifying step
one where $SU_{1}(4)$ was broken. Consider the block diagonal
decomposition of $SU_{1}(4)$ into
$SU_{1L}(2) \times SU_{1R}(2).$ The representations
$(2_{a}\alpha ,4_{1})$ and $(2_{a}\alpha ^{-1},4_{1})$ then decompose as

$(2_{a}\alpha ,4_{1})\rightarrow (2_{a}\alpha ,2,1)+(2_{a}\alpha ,1,2)$
and $(2_{a}\alpha ^{-1},4_{1})\rightarrow (2_{a}\alpha
^{-1},,2,1)+(2_{a}\alpha ^{-1},1,2)$. Now if we give $VEVs$ of equal
magnitude to the $(2_{a}\alpha ,,2,1),$ $a=1,2,3$, and equal magnitude
$VEVs$ to the $(2_{a}\alpha ^{-1},1,2)$ $a=1,2,3,$ we break
$SU_{1L}(2) \times \prod\limits_{a=1}^{3}SU(2_{a}\alpha )$ to a
single $SU_{L}(2)$ and we break
$SU_{1R}(2) \times \prod\limits_{a=1}^{3}SU(2_{a}\alpha )$ to a
single $SU_{R}(2).$  Finally, $VEVs$ for $(2_{4}\alpha ,2,1)\ $and
$(2_{4}\alpha ,1,2)$ as well as $(2_{4}\alpha ^{-1},2,1)\ $and
$(2_{4}\alpha ^{-1},1,2)$ insures that both $SU(2_{4}\alpha )$ and
$SU(2_{4}\alpha ^{-1})$ are broken and that only three families remain
chiral. The final set of chiral fermions is then
$3[(2,1,4)+(1,2,\bar{4})]$ with gauge symmetry
$SU_{L}(2) \times SU_{R}(2) \times SU_{D}(4).$

 To achieve the final reduction to the standard model, an adjoint VEV\
from
($\overline{4}_{\alpha }$,4$_{\alpha ^{2}})$ and/or
(4$_{\alpha }$,$\overline{4}_{\alpha ^{2}})$
is used to break $SU_{D}(4)$ to the
$SU(3)\times U(1),$ and a right handed doublet is used to break
$SU_{R}(2).$

While this completes our analysis of symmetry breaking, it is worthwhile
noting the degree of constraint imposed on the symmetry and particle
content of a model as the number of irreps $N_{R}$ of the discrete group
$\Gamma $ associated with the choice of orbifold changes. The number of
guage groups grows linearly in $N_{R}$, the number of scalar irreps
grows roughly quadratically with $N_{R}$, and the chiral fermion content
is highly $\Gamma $ dependent. If we require the minimal $\Gamma $ that
is large enough for the model generated to contain the fermions of the
standard model and have sufficient scalars to break the symmetry to that
of the standard model, then $\Gamma = Q \times Z_{3}$ appears to
be that minimal choice\cite{FK2}.

Although a decade ago the chances of testing string theory seemed at
best remote, recent progress has given us hope that such tests may
indeed be possible in AdS/CFTs. The madel provided here demonstrates the
standard model can be accomodated in these theories and suggests the
possibility of a rich spectrum of new physics just around the TeV
corner.

\bigskip
\bigskip
\bigskip

TWK thanks the Department of
Physics and Astronomy at UNC Chapel Hill for hospitality
while this work was in progress. This work was supported in part by the
US Department of Energy under
Grants No. DE-FG02-97ER-41036 and No. DE-FG05-85ER40226.

\bigskip
\bigskip
\bigskip
\bigskip

\newpage

.
\bigskip
\bigskip
\bigskip
\bigskip
\bigskip
\bigskip
\bigskip
\bigskip
\bigskip

{\bf Figure 1}

\bigskip
\bigskip
\bigskip
\bigskip

Quiver diagram for the chiral fermions in the $S^{5}/(Q\times Z_{3})$
orbifold. The arrows continue around the diagram, focusing on $4$'s and
diverging on doublets. Arrows pointing toward $4$'s give $(4,2)$ type
terms, while those pointing away give $(\overline{4},2)$'s.

\end{document}